\documentclass[prb,letterpaper,twocolumn,showpacs]{revtex4}

\usepackage{amsbsy,amssymb,amsmath,bm}
\usepackage{graphicx,color,epsfig,rotate}

\begin{document}

\topmargin -15mm
\preprint{APS/123-QED}

\title{Anisotropic phase diagram of the frustrated spin dimer compound Ba$_3$Mn$_2$O$_8$}

\author{E. C. Samulon$^1$, K. A. Al-Hassanieh$^2$, Y.-J. Jo$^3$, M. C. Shapiro$^1$, L. Balicas$^3$, C. D. Batista$^2$, I. R. Fisher$^1$}

\affiliation{$^1$Geballe Laboratory for Advanced Materials and Department of Applied Physics, Stanford University, Stanford, California 94305, USA}

\affiliation{$^2$Theoretical Division, Los Alamos National Laboratory, Los Alamos, New Mexico 87545, USA}

\affiliation{$^3$National High Magnetic Field Laboratory, Florida State University, Tallahassee, Florida 32306, USA}

\begin{abstract}

Heat capacity and magnetic torque measurements are used to probe the anisotropic temperature-field phase diagram of the frustrated spin dimer compound Ba$_3$Mn$_2$O$_8$ in the field range from 0T to 18T.  For fields oriented along the $c$ axis a single magnetically ordered phase is found in this field range, whereas for fields oriented along the $a$ axis two distinct phases are observed.  The present measurements reveal a surprising non-monotonic evolution of the phase diagram as the magnetic field is rotated in the [001]-[100] plane.  The angle dependence of the critical field ($H_{c1}$) that marks the closing of the spin gap can be quantitatively accounted for using a minimal spin Hamiltonian comprising superexchange between nearest and next nearest Mn ions, the Zeeman energy and single ion anisotropy.  This Hamiltonian also predicts a non-monotonic evolution of the transition between the two ordered states as the field is rotated in the $a$-$c$ plane. However, the observed effect is found to be significantly larger in magnitude, implying that either this minimal spin Hamiltonian is incomplete or that the magnetically ordered states have a slightly different structure than previously proposed.

\end{abstract}


\maketitle

\section{Introduction}

\begin{figure}
\includegraphics[width=8cm]{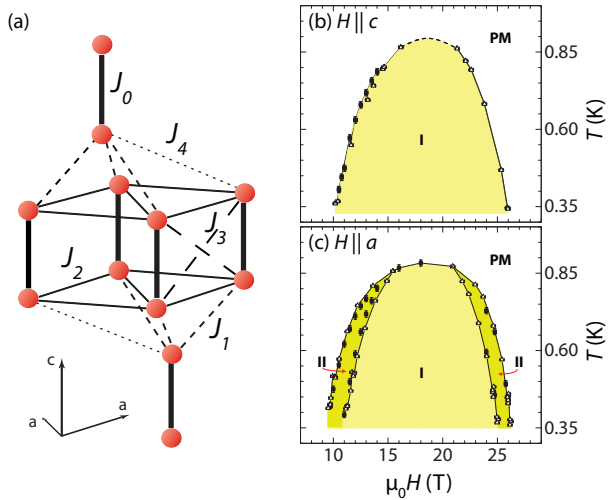}
\caption{(Color online) (a) Schematic diagram illustrating the exchange interactions between Mn ions (red spheres) in Ba$_3$Mn$_2$O$_8$ (Ba and O ions are not shown).  Values for the exchange constants are given in the main text   (b, c) Phase diagram of singlet-triplet ordered states for fields aligned along the $c$ and $a$ axes, respectively, taken from our previous work\cite{Samulon_2008}.  I, II and PM mark phase I, phase II and the paramagnetic phase as described in the main text.  Solid and open symbols represent data points obtained via heat capacity and MCE, respectively.}
\label{Cryst}
\end{figure}

Ba$_3$Mn$_2$O$_8$ is a novel layered spin-dimer compound, comprising magnetic dimers of Mn$^{5+}$ ions arranged on triangular planes\cite{Weller_1999}.  The Mn ions occupy equivalent sites with a distorted tetrahedral coordination, resulting in a quenched orbital angular momentum and a total spin $\mathbf{S}=1$ \cite{Uchida_2002}.  Antiferromagnetic intradimer exchange leads to a singlet ground state with excited triplet and quintuplet states. Weaker interdimer exchange leads to dispersion of these excitations, as previously revealed by inelastic neutron scattering \cite{Stone_2008}.  The rhombohedral R$\bar{3}$m crystal structure of Ba$_3$Mn$_2$O$_8$ comprises staggered hexagonal planes (Fig. \ref{Cryst}(a)), leading to geometric frustration both within individual planes, and also between adjacent planes. Magnetic fields can be used to close the spin gap, and the competition between interdimer interaction on this highly frustrated lattice, and the uniaxial single ion anisotropy associated with the $\mathbf{S}=1$ ions, leads to a very complex phase diagram with at least three distinct ordered states \cite{Samulon_2008, Samulon_2009}.

Despite the complexity of the phase diagram of Ba$_3$Mn$_2$O$_8$, we have previously shown that for fields oriented along the principal crystalline axes the high-field behavior of this material can be described by a remarkably simple spin Hamiltonian, including terms representing superexchange between nearest and next-nearest Mn ions within and between planes, the Zeeman energy and uniaxial single ion anisotropy \cite{Samulon_2008}.  These terms have been determined through a combination of inelastic neutron scattering (INS) and electron paramagnetic resonance (EPR).  INS studies revealed the dominant exchange within a dimer as $J_0$ = 1.642(3) meV; the nearest and next nearest out-of-plane exchanges, $J_1$ = 0.118(2) meV and $J_4$ = 0.037(2) meV; and the dominant in-plane exchanges $J_2 - J_3$ = 0.1136(7) meV (Fig. \ref{Cryst}) \cite{Stone_2008}.  EPR measurements in the diluted compound Ba$_3$(V$_{1-x}$Mn$_x$)$_2$O$_8$ revealed a nearly isotropic $g$-tensor, with $g_{aa}$ = 1.96 and $g_{cc}$ = 1.97, and an easy axis single ion anisotropy $D$ = -0.024 meV \cite{Whitmore_1993}.  Similar measurements in the pure Ba$_3$Mn$_2$O$_8$ compound revealed a zero field splitting of the triplet states characterized by $D$ =- 0.032 meV \cite{Hill_2007}, the difference reflecting the additional effect of intradimer dipolar coupling in the undiluted compound.

At low temperatures, the field dependence of the magnetization of Ba$_3$Mn$_2$O$_8$ reveals two regions of linearly increasing moment as first the $S^z=1$ triplet (between $H_{c1}=8.7$ T and $H_{c2}=26.5$ T) and then $S^z=2$ quintuplet (between $H_{c3}=32.5$ T and $H_{c4}=47.9$ T) states are polarized, separated by a plateau at half the saturation magnetization \cite{Uchida_2002, Samulon_2009}.  Here we concentrate solely on the first region of increasing magnetization, which can be accessed by moderate laboratory fields.  In this field range the ground state can be approximated as a coherent mixture of singlet and $S^z$=1 triplet states. Previous measurements revealed a distinct anisotropy in the phase diagram depending on the field direction \cite{Samulon_2008}.  For fields applied along the $c$ axis, a single ordered state was found in this regime (fig. \ref{Cryst}(b)).  Analysis of the minimal spin Hamiltonian suggests that this state is an incommensurate XY antiferromagnet with wavevector shifted slightly away from the 120$^{\circ}$ structure favored for a 2D triangular layer\cite{Samulon_2008, Diep_2005}.  NMR measurements appear to confirm this conclusion\cite{Suh_2009}, but to date the magnetic structure has not been explicitly solved. Alternatively, fields applied perpendicular to the $c$ axis revealed \textit{two} ordered states in this field regime (Fig. \ref{Cryst}(c)).  Analysis of the minimal spin Hamiltonian for this field direction yields two distinct incommensurate modulated states - an Ising phase stabilized closed to $H_{c1}$ and $H_{c2}$, as well as a canted XY state in between.  The modulation in these two phases is stabilized by the uniaxial anisotropy associated with the zero field splitting of the triplets.  Similar to the situation for fields oriented along the $c$ direction, the exact form of the magnetic structure has not been explicitly solved, and it remains to be seen whether the minimal spin Hamiltonian that has been used to describe this system so far is complete.

To this end, careful measurements of the phase diagram of Ba$_3$Mn$_2$O$_8$ for fields at intermediate angles between the $a$ and $c$ axes have the potential to determine the presence or absence of additional terms.  In this paper we present results of heat capacity and torque magnetometry measurements revealing how the two distinct ordered states for fields perpendicular to the $c$ axis evolve into a single phase for fields along the $c$ axis.  Through analysis of the previously established minimal spin Hamiltonian we can quantitatively account for the angular dependence of $H_{c1}$ solely via consideration of the triplet dispersion.  However, the same analysis, incorporating the predicted magnetic structures, fails to quantitatively account for the angular dependence of the transition between the two ordered states.  We discuss the implications of this observation.

\section{Experimental Methods}

Single crystals of Ba$_3$Mn$_2$O$_8$ were grown from a NaOH flux according to our previously published procedure\cite{Samulon_2008}.  Heat capacity ($C_p$) studies were performed on a Quantum Design physical properties measurement system (PPMS) using standard thermal relaxation-time calorimetry. These measurements were performed in fields up to 14T and temperatures down to 0.35 K.  The sample was mounted on angled brackets made from oxygen-free high conductivity copper and the field was oriented in the [100]-[001] plane.

Cantilever torque magnetometry experiments were performed at the National High Magnetic Field Laboratory (NHMFL) in a superconducting magnet for fields up to 18T in a dilution refrigerator.  A crystal was mounted on one face of a capacitance cantilever which was attached to a rigid plate rotatable about an axis parallel to the torque axis and perpendicular to the magnetic field.  The magnetic field was aligned away from the principal crystalline axes yielding a finite torque.

\section{Results}

\begin{figure}
\includegraphics[width=7cm]{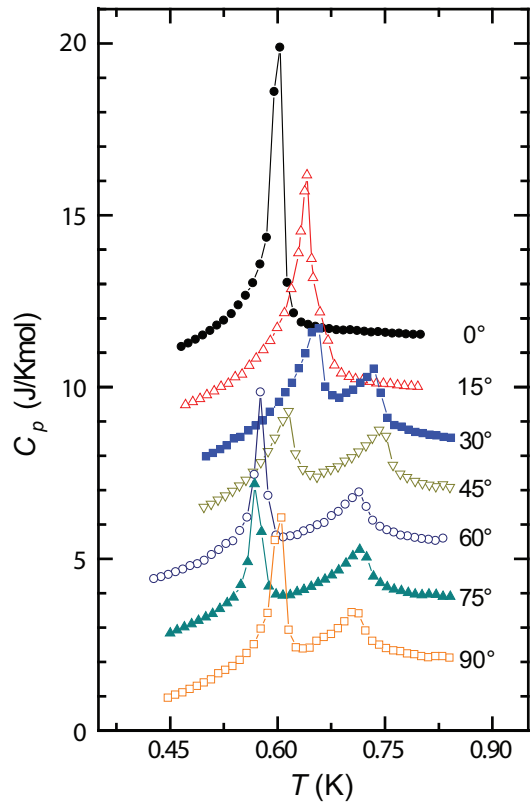}
\caption{(Color Online) Representative heat capacity data taken at 12T for for fields in the $a$-$c$ plane.  Labels indicate the angle between the field and the $c$ axis.  Successive data sets are offset vertically by 1.6 J/molK for clarity.}
\label{CpRef}
\end{figure}

Representative heat capacity measurements, taken at 12 T for several angles, are shown in Figure \ref{CpRef}.  These data show a single peak for fields aligned along the $c$ axis, a peak and a shoulder for fields 15$^{\circ}$ from the $c$ axis and two peaks for larger angles.  Significantly, comparison of the data at 75$^{\circ}$ and 90$^{\circ}$ degrees shows that the 75$^{\circ}$ data has both a slightly higher critical temperature between the paramagnetic phase and phase II ($T_{c_{II}}$) and also a substantially lower critical temperature between phase II and phase I ($T_{c_{I}}$) than the 90$^{\circ}$ data.

\begin{figure}
\includegraphics[width=8.5cm]{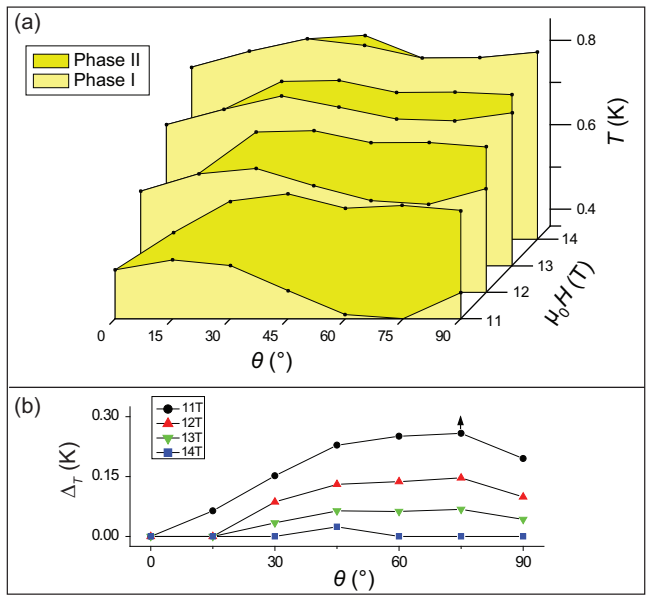}
\caption{(Color online) (a) Phase diagram showing the transitions between the paramagnetic state and phase II ($T_{II}$), and between phase II and phase I ($T_I$), as a function of temperature and angle in the [100]-[001] plane for various fields, where $\theta$ indicates the angle between the field and the $c$ axis. (b) Width of phase II, $\Delta_T = T_{c_{II}}-T_{c_{I}}$, as a function of angle for 11T, 12T, 13T and 14T (black circles, red up triangles, green down triangles, and blue squares, respectively.)}
\label{Cp}
\end{figure}

The phase diagram derived from the complete set of $C_p$ measurements, shown in Figure \ref{Cp}(a), reveals the evolution as a function of angle of the two distinct singlet-triplet ordered states for fields in the [100]-[001] plane.  The data show a single transition for all fields for $H \| c$ and two transitions for $H \| a$.  The extent in temperature of phase II, $\Delta_T=T_{c_{II}}-T_{c_{I}}$, is shown in Figure \ref{Cp}(b).  $\Delta_T$ increases as a function of angle as the field is rotated away from the $c$ axis, reaches a maximum at 75$^{\circ}$, and decreases at the $a$ axis (90$^{\circ}$).  For example, for a field of 11T, $\Delta_T$ is $\sim$ 0.07 K larger at 75$^{\circ}$ than at 90$^{\circ}$.

\begin{figure}
\includegraphics[width=8.5cm]{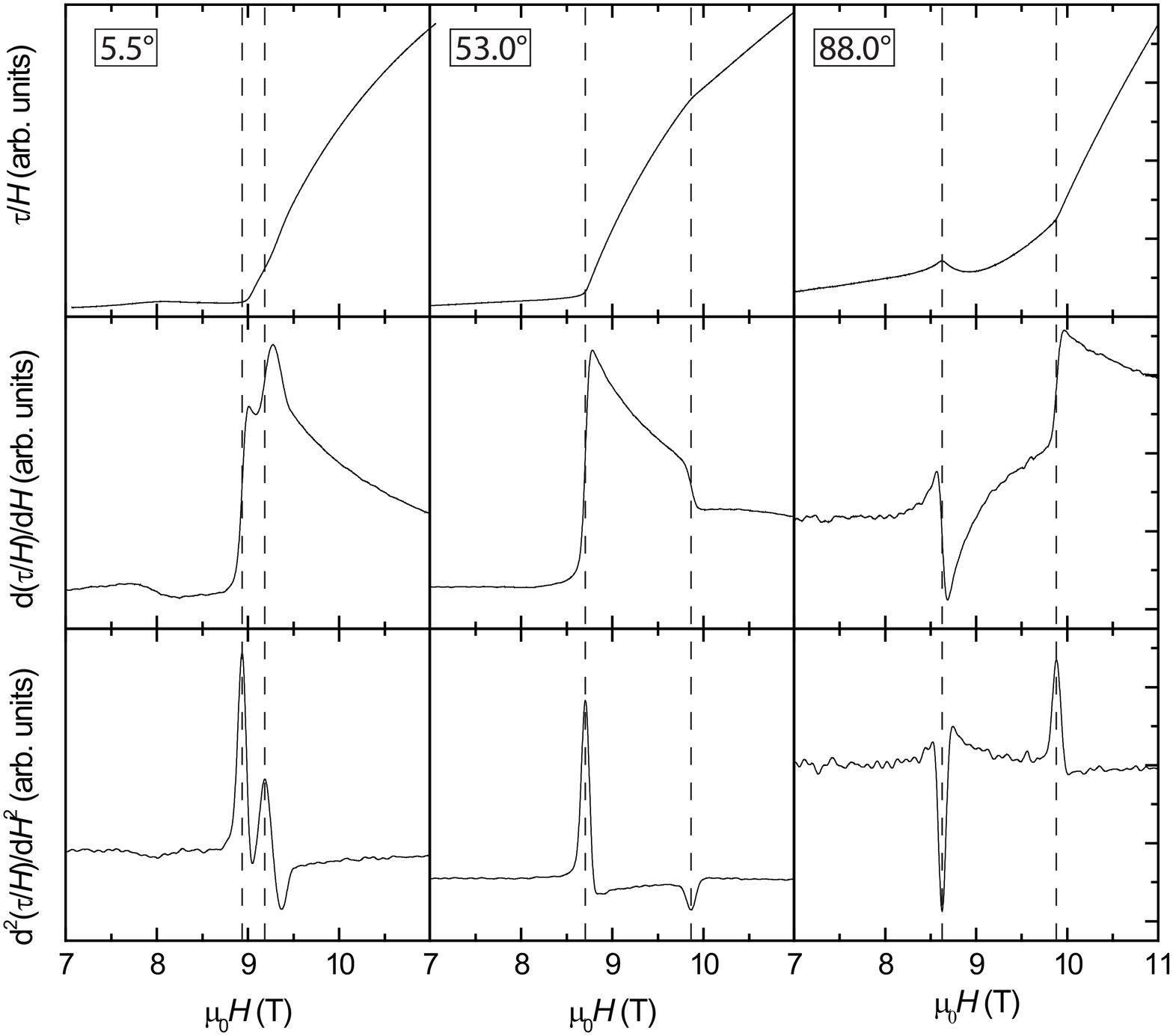}
\caption{Torque scaled by field, and its first two derivatives with respect to field, plotted versus field for representative angles at 5.5$^{\circ}$, 55.0$^{\circ}$ and 88.0$^{\circ}$ with respect to the nearest approach to the $c$ axis (see main text).  The critical fields were estimated from peaks in the 2$^{nd}$ derivative as shown by the dashed lines.}
\label{TScans}
\end{figure}

Torque magnetometry measurements, taken at 25 mK, are shown in Figure \ref{TScans} for three representative angles.  The sample was inclined slightly so that the field did not exactly rotate within the $a$-$c$ plane such that a finite torque was generated for all angles studied. Angles are quoted in terms of the angular position with respect to the closest approach to the $c$ axis, but it is important to note that the field was never less than $\sim 10^{\circ}$ from the $c$ axis. Consequently two phase transitions are observed for all angles studied.  Critical fields, marking the transition between the paramagnetic phase and phase II ($H_{c1}$), and between phase II and phase I ($H_{II-I}$), were determined from maxima and minima in the second field derivative of the torque divided by field \cite{Samulon_2008} and are marked with dashed vertical lines.  The sign of the peak in the second derivative changed with the evolution of angle, reflecting a change in anisotropy for the two different phases.  This leads to minor discontinuities in the determination of the phase boundary (dashed lines in Fig. \ref{Torq}).

Similar to the phase diagram obtained from heat capacity measurements (Fig. \ref{Cp}), the phase diagram obtained from torque measurements at 25mK reveals a non-monotonic angle dependence (Fig. \ref{Torq}).  The maximum value of $H_{II-I}$ occurs between 65-75$^{\circ}$ from the closest approach to the $c$ axis.  This is in agreement with the heat capacity data, for which the smallest $T_{c_I}$ occurs at 75$^{\circ}$ from the $c$ axis.  Additionally, the field extent of phase II, $\Delta_H = H_{II-I}-H_{c1}$, is largest at 75$^{\circ}$, and decreases by $\sim$0.05 T from the maximum at the highest angles.

\section{Discussion}

The previously established minimal spin Hamiltonian for arbitrarily oriented field direction in Ba$_3$Mn$_2$O$_8$ is:
\begin{eqnarray}
\mathcal{H} & = & \sum_{i, j, \mu, \nu} \frac{J_{i \mu j \nu}}{2}\textbf{S}_{i\mu}\cdot \textbf{S}_{j\nu}
+ D\sum_{i, \mu}\left(S^{z}_{i \mu} \cos{\theta} - S^{x}_{i \mu} \sin{\theta} \right)^2
\nonumber\\
&& -  \mu_B H \sum_{i \mu \alpha \beta}\left({\tilde g}_{zz} S^{z}_{i \mu} + {\tilde g}_{xz} S^{x}_{i \mu}\right),
\label{Ham}
\end{eqnarray}
where ${\tilde g}_{zz}=g_{aa} \sin^2{\theta}+g_{cc} \cos^2{\theta}$, ${\tilde g}_{xz}=(g_{cc}-g_{aa}) \sin{\theta} \cos{\theta}$, $g_{\alpha \beta}$ is the diagonal gyromagnetic tensor with components $g_{cc}$, $g_{aa}=g_{bb}$, and
$\theta$ is the angle between the applied field and the $c$-axis. The quantization $z$ axis is set along the field direction.  Here $i$, $j$ designate the dimer coordinates, $\alpha,\beta=\{x,y,z\}$, $\mu,\nu=\{1,2\}$ denote each of the two S=1 spins in  each dimer. The various exchange constants are shown in Fig. \ref{Cryst}(a) and are defined as follows: the exchange within a dimer is $J_0=J_{i, 1, i, 2}$; the dominant out-of-plane exchange is $J_1 = J_{i, 2, j, 1}$ for $i,j$ nearest neighbor dimers between planes; the dominant in-plane exchanges between dimers is $J_2 = J_{i, \mu, j, \mu}$ and $J_3 = J_{i, \mu, j, \nu}$ for $i,j$ in plane nearest neighbor dimers and $\mu \neq \nu$; and finally the second largest out-of-plane exchange is $J_4 = J_{i, 2, j, 1}$ for $i,j$ next nearest neighbor dimers between planes.

\begin{figure}
\includegraphics[width=8.5cm]{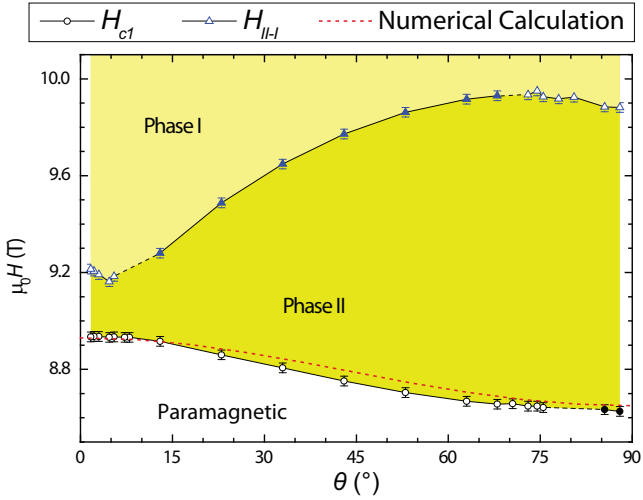}
\caption{(Color online) Phase diagram at 25 mK determined from torque magnetometry measurements.  Circles (triangles) mark transitions between the disordered phase and phase II (phase II and phase I).  Open (closed) symbols signify that the transition was determined from a peak (trough) in the second derivative.  Angles are measured relative to the closest approach to the $c$ axis as described in the main text.  Red dotted line shows the calculated transition between the paramagnetic and ordered phases as described in the main text.}
\label{Torq}
\end{figure}

Using this spin Hamiltonian and the measured values of $J_0$-$J_4$ and $D$ given in the introduction, we have previously been able to quantitatively account for the observed critical fields and magnetization of the ordered states of Ba$_3$Mn$_2$O$_8$ for fields oriented along the principal axes \cite{Samulon_2008, Samulon_2009, Suh_2009}.  The calculation is based on a generalized spin-wave approach in which we only keep the singlet and the three triplet states of each dimer \cite{Khaled2010}. The critical field $H_{c1}(\theta)$ corresponds to the value for which the energy of lowest energy triplet mode becomes equal to zero. The softening of this triplet mode signals the onset of the magnetic instability towards an ordered state (phase II). For field directions along the principal axes, it is possible to obtain simple analytical expressions for the critical field. The expression for $H \| c$ is:
\begin{equation}
(g_{cc} \mu_B H_{c1})^2 = \left(J_0-\frac{D}{3}\right)^2 +\frac{8}{3}\left(J_0 - \frac{D}{3}\right)\mathcal{J}_{min},
\end{equation}
while for $H \perp c$ the expression is:
\begin{eqnarray}
(g_{aa} \mu_B H_{c1})^2 & = & \left(J_0+\frac{D}{6}\right)^2 +\frac{8}{3}\left(J_0 + \frac{D}{6}\right)\mathcal{J}_{min}
\nonumber\\
&& -  \frac{D^2}{4}-\frac{4}{3}|D||\mathcal{J}_{min}|
\end{eqnarray}
where $\mathcal{J}_{min}$ is the minimum of the interdimer exchange portion of the dispersion and is fully described in equation (3) of reference [\onlinecite{Stone_2008}].  The difference in the first two terms between these expressions stems from a change in the zero field splitting of an isolated dimer depending on quantization direction expressed in the reduced basis of dimer states. For the quantization axis along the $c$ axis, the zero field gap of an isolated dimer between the $S^z=1$ triplet and the singlet is $J_0 - D/3$, while for the quantization axis along the $a$ axis the equivalent zero field gap is $J_0+D/6$. The two additional terms of $H_{c1}$  for fields perpendicular to the $c$ axis arise from a second order process mixing singlets and triplets described in our previous work \cite{Samulon_2008}.  This state mixing causes the gap between the $S^z=1$ triplet and singlet states to close as $\sqrt{H-H_{c1}}$, as expected for an Ising-like QCP.   Evaluating these expressions using the values of the exchanges and single ion anisotropy described earlier yield values for the critical fields which are in good accord with the measured data.  Numerical calculation of $H_{c1}$ for arbitrary field orientations in the [001]-[100] plane yields the red dotted curve shown in Fig. \ref{Torq}.  The calculation was performed using the values of $D$ and the interdimer couplings given in the introduction while $J_0$ was allowed to vary leading to a fit value of 1.567 meV.  The calculated values agree well with the measured data up to the inherent uncertainty associated with the misalignment of the sample in the torque measurements described above.

The single-ion anisotropy  term of ${\cal{H}}$ is:
\begin{eqnarray}
D\bigg[ \left(S^z\right)^2\cos^2\left(\theta\right) + \left(S^x\right)^2\sin^2\left(\theta\right)
\nonumber\\
- \left(S^zS^x + S^xS^z\right) \cos\left(\theta\right) \sin\left(\theta\right) \bigg].
\label{Single}
\end{eqnarray}
The last term of eq. (\ref{Single}) is zero for fields along the $a$ and $c$ axes, but adds a small contribution for intermediate angles.  In particular, this term grows linearly in small deviations of $\theta$ from $\pi/2$, $\delta \theta =\pi/2-\theta$, while the other two terms vary quadratically in $\delta \theta$.  Thus, the last term of Eq. (\ref{Single}) determines the shape of the boundary between phase I and phase II slightly away from $\theta = \pi/2$, and could be responsible for the striking non-monotonic behavior observed for $H_{II-I}$ in Fig. \ref{Torq} as we describe in greater detail below.

To understand the effect that the last term of eq. (\ref{Single}) has on the ground state, it is convenient to analyze the effective low-energy Hamiltonian, ${\cal H}_{\rm eff}$, introduced in ref. [\onlinecite{Samulon_2008}]. The low-energy Hamiltonian results from projecting the original Hamiltonian ${\cal H}$ onto the low-energy subspace generated by the singlet and the $S^z=1$ triplet states of each dimer. This two-level Hilbert space is described by a local pseudo-spin $\frac{1}{2}$ in each dimer, ${\bf s}_i$, such that $s^z_i=\frac{1}{2}$ if the dimer $i$ is in the $S^z=1$ triplet state and  $s^z_i=-\frac{1}{2}$ if it is in the singlet state. In our earlier work we provided the expression of ${\cal H}_{\rm eff}$ for $\theta=0$ and $\theta=\pi/2$ \cite{Samulon_2008}. In particular, for $\theta=\pi/2$, we showed that the second term of  Eq. (\ref{Single}) generates an effective exchange anisotropy that is responsible for the emergence of phase II. According to our analysis, this exchange anisotropy favors an Ising-like phase in which the transverse spin components (perpendicular to the applied field) are aligned along the easy $c$-axis. In contrast, phase I is an elliptical spiral phase in which the transverse spin components of adjacent dimers rotate around the field axis.

The effect the last term of Eq. (\ref{Single}) has on ${\cal H}_{\rm eff}$ at intermediate angles can be determined using second order degenerate perturbation theory.  Such an analysis yields an effective Dzyaloshinskii-Moriya (DM) interaction between dimers on adjacent bilayers connected by the $J_1$ and $J_4$ exchange constants:
\begin{equation}
\sum_{\langle \langle i \rightarrow j \rangle \rangle} {\bf {\tilde D}}_1 \cdot {\bf s}_i \times {\bf s}_j
+ \sum_{\langle \langle i \rightarrow j \rangle \rangle'} {\bf {\tilde D}}_4 \cdot {\bf s}_i \times {\bf s}_j,
\label{efff2}
\end{equation}
with  ${\bf {\tilde D}}_l = {\tilde D}_l {\bf {\hat y}}$, ${\tilde D}_l = {\cal O}(DJ_l/J_0)$ and $l=1,4$. The arrow indicates how the bonds $\langle \langle i \rightarrow j \rangle \rangle$ are oriented ($i$ always denotes the dimer in the lower bilayer). Microscopically, this process turns one singlet into an $S^z=1$ triplet or vice versa.

\begin{figure}[!htb]
\includegraphics[angle=0,width=8.0cm]{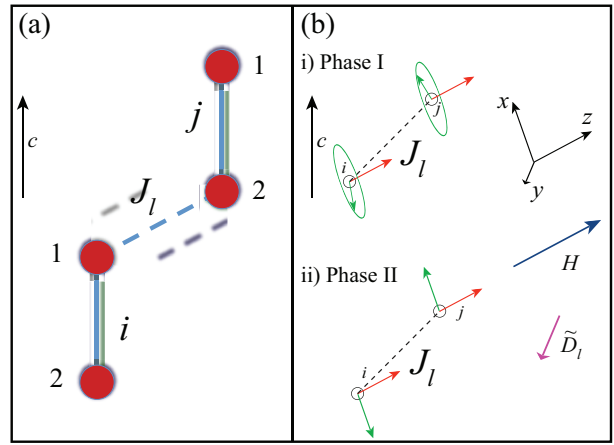}
\caption{(a) Schematic diagram illustrating the two dimers involved in processes of order $ J_l D/J_0$ that leads to Eq. (\ref{efff2}).  The $c$ axis is vertical, and the two dimers belong to adjacent planes.  $J_l$ represents either $J_1$ or $J_4$. (b) (i), (ii) Schematic diagram for two pseudo-spins on adjacent dimers connected by the $J_l$ interaction for arbitrary field direction in the $a$-$c$ plane.  Red arrows show uniform moment, while green arrows show ordered moment.  Field direction along $z$ axis, ${\bf {\tilde D}}_l$ along $y$ axis, and $c$ axis vertical.}
\label{twod}
\end{figure}

This effective DM coupling between pseudo-spins results from two important symmetry considerations. First, the singlet and the triplet states of a given dimer have opposite parity under exchange of the two sites of the dimer: $1 \leftrightarrow 2$. Second, the $J_1$ and $J_4$ terms of ${\cal H}$ are invariant under the inversion symmetry transformation around the center of the corresponding bonds: $1 \leftrightarrow 2$ and $i \leftrightarrow j$ (see Fig. \ref{twod}(a)). This implies that a DM term is allowed between pseudo-spins connected by the $J_1$ and $J_4$ exchanges.  In contrast, the effective DM interaction cannot occur for pairs of dimers within a plane because the $J_2$ and $J_3$ terms of ${\cal H}$ are invariant under the symmetry transformation $i \leftrightarrow j$, while the effective DM interaction changes sign under such transformation.  These symmetries are fundamentally equivalent to the selection rules governing the DM vector in the effective lattice, where each dimer constitutes a single, unique site.  In such a lattice, there is a center of inversion symmetry at the midpoint of the effective $J_2$ and $J_3$ exchange, precluding a DM term by the normal selection rules, while no such inversion symmetry exists at the midpoints of the effective $J_1$ and $J_4$ exchanges.  This is contrast to the symmetries of the $\textit{real}$ lattice, where there are inversion symmetries at the midpoint of the $J_1$ and $J_4$ exchanges but none at the midpoint of the $J_2$ and $J_3$ exchanges.

This effective DM interaction between dimers on adjacent layers is frustrated in both ordered states at a mean field level (the mean value of Eq. (\ref{efff2}) is zero for the semi-classical states associated with phases I and II). Therefore, the small contribution of the effective DM interaction to the ground state energy must be produced by quantum fluctuations.  Because the DM vectors point along the $y$ direction (perpendicular to the applied field and to the easy $c$ axis) this contribution term will favor phase II, for which the pseudo-spins only have $x$ and $z$ components (Fig. \ref{twod}(b)(ii)), as opposed to phase I for which the pseudo-spins have an additional third component along the $y$ direction (Fig. \ref{twod}(b)(i)) \cite{Samulon_2008}.  Thus this contribution strengthens phase II relative to phase I near $\theta=\pi/2$ and leads to a small non-monotonic behavior of the $H_{II-I}$ curve (see Fig. \ref{Torq}).

Although this simple analysis captures the qualitative non-monotonic behavior of the $H_{II-I}$ curve, it cannot account for the magnitude of the observed effect.  The amplitude ${\cal D}(\theta)=D \sin{\theta} \cos{\theta}$ of the effective DM interaction is of order 100 mK for ${\theta} \simeq 75^{\circ}$. Because the interaction mixes the singlet and $S^z=1$ triplet dimer states, the mean value of the DM term is less than $\sqrt{m} {\cal D}(\theta)$ for any state with magnetization $m$ (the mean density of $S^z=1$ triplets).  Noting that $m \simeq 0.06$ at $H_{II-I}$, the upper bound on the DM term of $\simeq 25$ mK is the order of magnitude of the observed non-monotonic effect of 5-10 mK in the $H_{II-I}$ curve. Given that ${\cal D}(\theta)$ is much weaker than the dominant terms of ${\cal H}$, it is  clear that the effective DM term can only explain the magnitude of the non-monotonic effect if it gives a first order contribution to the energy of Phase II. However, as established above for the proposed ordered states, the effective DM interaction contributes via a second order correction and must therefore be considerably smaller.  This leaves us with two possibilities: a) The magnetic structure of Phase II is different from the simple Ising phase proposed in Ref. [\onlinecite{Samulon_2008}] in such a way that the mean value of the effective DM term is non-zero, or b) The non-monotonic effect is caused by a term that has not been included in ${\cal H}$.  At present it is impossible to distinguish between these possibilities, but ongoing efforts to experimentally determine the magnetic structure have the potential to directly address option (a), while EPR experiments should, at least in principle, be able to determine the energy scale of additional interactions not considered in the minimal spin Hamiltonian (eq. (\ref{Ham})).

\section{Conclusion}

In summary, via heat capacity and torque magnetometry measurements we have established the angular dependence of the phase boundary for singlet-triplet ordered states of Ba$_3$Mn$_2$O$_8$.  The data reveal a striking non-monotonicity of the phase boundary as the field is rotated between the principal axes.   The angle-dependence of $H_{c1}$ can be quantitatively understood in terms of the original minimal spin Hamiltonian that we had proposed for this material.  This quantity does not depend on details of the magnetically ordered states but only on the minimum of the triplet dispersion. However, the observed non-monotonicity in $H_{I-II}$ is at least an order of magnitude larger than anticipated based on this model and assuming the magnetic structures previously proposed.  This indicates that a complete theoretical description of this material requires either subtle changes in the proposed magnetic ordered structures or an additional low-energy term in the Hamiltonian.

\section{Acknowledgements}

Work at Stanford University is supported by the Division of Materials Research, National Science Foundation under Grant No. DMR-0705087.  LB is supported by DOE-BES and the NHMFL-UCGP program.  A portion of this work was performed at the National High Magnetic Field Laboratory, which is supported by the National Science Foundation under Cooperative Agreement No. DMR-0084173, by the state of Florida, and the Department of Energy.


\begin{thebibliography}{1}
\bibitem{g_ani}  While the $g$ factor will also play a role in the angle dependence of phase II, it is nearly isotropic and will vary monotonically as a function of angle.
\end{thebibliography}
\end{document}